\documentclass[10pt,a4paper]{article}

\usepackage{a4}
\usepackage{amssymb}
\usepackage{bbm}
\usepackage{mathrsfs}
\usepackage{eucal}
\usepackage{fancyhdr}

\newcommand{\one}{\mathbbm{1}}
\newcommand{\V}{\tilde{V}}
\newcommand{\bra}[1]{\left\langle\hspace{0.5pt}#1\right.\left.\!\!\right|}
\newcommand{\ket}[1]{\left|\hspace{0.5pt}#1\right.\left.\!\!\right\rangle}

\begin{document}
\begin{titlepage}
\setcounter{page}{1}
\renewcommand{\thefootnote}{\fnsymbol{footnote}}

\begin{flushright}
  HD-THEP-03-30 \\
  \ \\
\end{flushright}

\vspace{5 mm}

\begin{center}
{\Large\bf Resummed effective action in the world-line formalism}

\vspace{10 mm}

{\bf Thomas Auer\footnote{e-mail: thomas.t.auer@web.de}},
{\bf Michael G. Schmidt\footnote{e-mail: m.g.schmidt@thphys.uni-heidelberg.de}}
\textsc{and}
{\bf Claus Zahlten\footnote{e-mail: zahlten@thphys.uni-heidelberg.de}}

\vspace{15mm}

{\it Institut f\"ur Theoretische Physik\\
     Universit\"at Heidelberg\\
     Philosophenweg 16, D-69120 Heidelberg, Germany}
\end{center}

\vspace{18mm}

\begin{abstract}
\noindent
Using the world-line method we resum the scalar one-loop effective action.
This is based on an exact expression for the one-loop action obtained
for a background potential and a Taylor expansion of the potential 
up to quadratic order in x-space. We thus reproduce results of Masso 
and Rota very economically. An alternative resummation scheme is 
suggested using ``center of mass'' based loops which is equivalent
under the assumption of vanishing third and higher derivatives in the
Taylor expansion but leads to simplified expressions.
In an appendix some general issues concerning the relation between 
world-line integrals with fixed end points versus integrals with fixed 
center are clarified. 
We finally note that this method is also very valuable for gauge field 
effective actions where it is based on the Euler--Heisenberg type 
resummation.
\end{abstract}

\vfill

\end{titlepage}

\setcounter{footnote}{0}
\renewcommand{\theequation}{\thesection.\arabic{equation}}

\section{Introduction}
\setcounter{equation}{0}

In this paper we reconsider the one-loop contribution to the effective 
action for a scalar theory with self-interaction potential $V(\phi(x))$. 
Based on the momentum space method of ref.~\cite{Brown:bc} E.~Masso 
and F.~Rota have shown that the terms up to quadratic order in the 
Taylor expansion of the potential can be resummed 
exactly \cite{Masso:2001zh}.

In the framework of the world-line formalism 
\cite{Strassler:1992zr, Schmidt:rk, Schubert:Review} this fact is 
not surprising: world-line path integrals can be computed analytically 
when the action contains only terms 
of at most quadratic order in the world-line fields. 
In addition to that, in general the world-line formalism has some 
valuable advantages in calculating effective actions: the outer 
fields can be taken into account as a background in representation 
space easily which leads to very compact and elegant expressions.

Both of these facts suggest that the exact resummation of quadratic terms
performed in ref.~\cite{Masso:2001zh} should be possible
in a much more straight forward way using world-line methods.

The paper is organized as follows. In section \ref{first_WL_approach} we
calculate the resummed one-loop contribution to the effective action
using world-line methods. We confirm the result obtained in 
\cite{Masso:2001zh} and show that indeed within the world-line formalism
the calculation demands much less effort.
In section \ref{second_WL_approach} we suggest an alternative
resummation scheme using the freedom to choose a different organization 
of the paths to be integrated over inherent to the world-line formalism.
This results in a very efficient way of resumming terms up to second 
order and leads to simplified expressions. 
Section \ref{inverseMass} is devoted to the relation between the two
resummation schemes and to a short comparison of both 
results with the invers mass expansion of the effective 
action given in refs.~\cite{Fliegner:1994zc, Fliegner:1997rk}. 
Finally, in section \ref{conclusions} we show how the presented 
formalism can equally well be applied to the case of gauge fields and 
conclude our results. In the appendix we discuss some general issues
concerning the relation between world-line path integrals with 
fixed end points versus integrals with fixed ``center of mass''.

\section{Resummation of the effective action}
\label{first_WL_approach}

In {\em Euclidean} space time conventions the world-line representation of 
the one-loop contribution to the effective action reads 
(see e.g.~\cite{Schubert:Review})
\begin{equation}
\label{wpi}
\Gamma^{(1)} [\varphi] 
 =  \frac{1}{2} \int\limits_{0}^{\infty} \frac{dT}{T}\,e^{-m^{2}T} 
    \!\!
    \int d^D\!q \!\!\int\limits_{x(0)=q}^{x(T)=q} \!\!\!\!\mathcal{D}x\;
    \exp\left\{ -\int\limits_{0}^{T} \!\!d\tau 
    \left[ \frac{1}{4}\,\dot{x}^{2}(\tau)  +
    V^{\prime\prime}
    \!\left( \varphi(x) \right) \right] \right\}  
\end{equation}
We will now Taylor expand the self-interaction potential, absorb the terms 
up to the quadratic order into the free path integral and resum the one-loop 
contribution to the effective action by performing this path integral.
Higher terms can then be obtained by contracting the world-line fields with 
the appropriate Green's function.

Writing for brevity $\V(x) = V^{\prime\prime}(\varphi(x))$
and expanding around the common starting and end point $q_{\mu}$ of the 
closed paths $x_{\mu}(\tau)$ we have
\begin{eqnarray} \label{taylor}
\V(x(\tau))\! 
& = & \!e^{(x(\tau)-q)_{\mu} \partial_{\mu}} \left.\V(x)\right|_q \\
& = & \! \V(q) + \beta_{\mu} \left( x(\tau) - q \right)_{\mu} 
       + \frac{1}{4}\,\gamma_{\mu\nu}^{2} 
       \left( x(\tau) - q \right)_{\mu} 
       \left( x(\tau) - q \right)_{\nu} 
       + \V_3(x(\tau)-q) \nonumber
\end{eqnarray}
where we have introduced the abbreviations
\begin{eqnarray}
\beta_{ \mu} & = & \partial_{\mu}\!\left. \tilde{V}(x) \right|_{q}
\label{ak1} \\ 
\gamma^{2}_{\mu\nu} & = &  \left(\gamma^2 \right)_{\mu\nu} = 
\ 2 \,\partial_{\mu}\partial_{\nu}\! \left. \tilde{V}(x) \right|_{q}
\label{ak2} \\
\V_3(x(\tau)-q) & = & \sum_{n=3}^{\infty} \frac{1}{n!}
                      \left((x(\tau)-q)\cdot\partial\right)^n
                      \left. \tilde{V}(x) \right|_{q} \label{ak3}
\end{eqnarray}
in accordance with ref.~\cite{Masso:2001zh}.
Inserting the expansion (\ref{taylor}) into eq.~(\ref{wpi}), 
shifting the integration path according to
\begin{math}
  y_{\mu}(\tau) =  x_{\mu} (\tau) - q_{\mu} 
\end{math}
and performing a partial integration in the exponent
yields
\begin{eqnarray} \label{eq:GammaFirstShift}
\lefteqn{
\Gamma^{(1)}[\varphi]  =  
\frac{1}{2} \int\limits_{0}^{\infty} \frac{dT}{T}\,e^{-m^{2}T}
\!\!\int d^{D}q \;e^{-{\V(q)T}}} \\
& & \times\;
\int\limits_{y(0)=0}^{y(T)=0} \!\!\!\!\mathcal{D}y\;
\exp\Biggl\{ -\int\limits_{0}^{T}\!\! d\tau 
\biggl[
  -\frac{1}{2}\,y_{\mu}(\tau)
      \left( \frac{1}{2}\,\delta_{\mu\nu}\partial_{\tau}^{2}  
            -\frac{1}{2}\,\gamma_{\mu\nu}^{2}
      \right)   y_{\nu}(\tau) 
  + \,\beta_{\mu} y_{\mu}(\tau) + \V_3(y(\tau))
\biggr]
\Biggr\} \nonumber
\end{eqnarray}
To further evaluate this world-line path integral we want to get rid of the
linear term in the exponent. This is achieved by completing the square
with another shift in the integration path
\begin{equation}
\label{zerl}
  \tilde{y}_{\mu}(\tau) = y_{\mu}(\tau) - \eta_{\mu}(\tau)
\end{equation}
where
\begin{equation} \label{eta}
  \eta_{\mu}(\tau) 
  = \int\limits_{0}^{T} d\tau^{\prime}
    G_{\,\mu\nu}^{\scriptscriptstyle\mathrm{(00)}}
    (\tau,\tau^{\prime})\beta_{\nu}\;
  = \;\beta_{\nu} \int\limits_{0}^{T} d\tau^{\prime}
    G_{\,\mu\nu}^{\scriptscriptstyle\mathrm{(00)}}(\tau,\tau^{\prime})
\end{equation}
and $G^{\scriptscriptstyle\mathrm{(00)}}$ is the appropriate Green's 
function defined by the following three properties 
\begin{eqnarray}
\left(
   \frac{1}{2} \delta_{\mu\nu} \partial_{\tau}^{2} 
  -\frac{1}{2} \gamma_{\mu\nu}^{2} 
\right) G^{\scriptscriptstyle\mathrm{(00)}}_{\,\nu\rho}(\tau,\tau^{\prime})
& = & \delta_{\mu\rho}\,\delta(\tau - \tau^{\prime}) \label{green1} \\
G^{\scriptscriptstyle\mathrm{(00)}}_{\,\mu\nu} (0,\tau^{\prime}) 
& = & 0 \label{green2} \\[1.2ex]
G^{\scriptscriptstyle\mathrm{(00)}}_{\,\mu\nu} (T,\tau^{\prime}) 
& = & 0 \label{green3}
\end{eqnarray}
With $G^{\scriptscriptstyle\mathrm{(00)}}$ satisfying 
eqs.~(\ref{green1}) to (\ref{green3})
it is ensured that the shifted paths 
also run from $\tilde{y}(0)=0$ to $\tilde{y}(T)=0$ and that boundary 
terms vanish. The actual form of 
$G^{\scriptscriptstyle\mathrm{(00)}}(\tau,\tau^{\prime})$ will be 
calculated in a moment. For now we
can just rely on properties (\ref{green1}) to (\ref{green3}) to
perform the shift. This leads to
\begin{eqnarray} \label{EffAction:0-0}
\Gamma^{(1)}[\varphi] & = & 
\frac{1}{2} \int\limits_{0}^{\infty} \frac{dT}{T}\,e^{-m^{2}T} 
\!\!\int d^{D}q \;e^{-{\V(q)T}}
\exp\left\{
  -\frac{1}{2}\,\beta_{\mu}\beta_{\nu}\!
  \int\limits_0^T d\tau d\tau^{\prime} 
  G^{\scriptscriptstyle\mathrm{(00)}}_{\,\mu\nu}(\tau,\tau^{\prime})
\right\}
\nonumber \\
& & \!\!\!\!\!\!\!
\times
\int\limits_{\tilde{y}(0)=0}^{\tilde{y}(T)=0} \!\!\!\!\mathcal{D}\tilde{y}
\;\exp\Biggl\{ -\int\limits_{0}^{T} d\tau 
\biggl[
  -\frac{1}{2}\,\tilde{y}_{\mu}(\tau)
      \!\left( \frac{1}{2}\delta_{\mu\nu}\partial_{\tau}^{2}  
            -\frac{1}{2}\gamma_{\mu\nu}^{2}
      \right)\!   \tilde{y}_{\nu}(\tau)
      + \V_3(\tilde{y} + \eta)
\biggr]
\Biggr\} \qquad
\end{eqnarray}
Ignoring the terms of third and higher order in derivatives of the potential,
i.e.~ignoring $\V_3$, yields the resummed effective action we are looking 
for. The influence of these higher derivatives can be taken into account
by treating $\V_3$ as an ``interaction term'' and calculating the resulting 
contractions by means of the resummed Green's function 
$G^{\scriptscriptstyle\mathrm{(00)}}$.

What remains to complete the calculation is just to perform the quadratic
path integral in eq.~(\ref{EffAction:0-0}) and to determine the Green's
function $G^{\scriptscriptstyle\mathrm{(00)}}$. 
By the defining property (\ref{green1}) the Green's 
function inverts the operator
\begin{equation}
  K = \frac{1}{2}\left(\one\,\partial_{\tau}^{2} - \gamma^{2}\right)
\end{equation}
which has to be read 
on the subspace of functions with $x(0) = x(T) = 0$ and
$x \in \mathbbm{R}^D$. 
We  now introduce a complete set of one-dimensional
normalized eigenfunctions of the operator $\partial_{\tau}^2$ that obey
the boundary condition $f(0)=f(T)=0$
\begin{equation} \label{eigenfuncs}
  \left\langle\hspace{0.5pt}\tau\hspace{0.5pt}|
              \hspace{0.5pt}f_n\hspace{0.5pt}\right\rangle = 
  f_n(\tau) = \frac{-i}{\sqrt{2T}}
              \left(
                e^{+i\pi n \tau/T} - e^{-i\pi n \tau/T}
              \right)
  \qquad;\ n \in \mathbbm{N}\setminus\{0\}
\end{equation}
In addition we introduce a set of normalized 
eigenvectors $\ket{v_1},\dots,\ket{v_D}$ of the matrix $\gamma^2$.
The corresponding eigenvalues $\partial_{\tau}^2 f_n(\tau) = c_n f_n(\tau)$
are given by
\begin{equation} \label{eigenvals}
  c_n = - \frac{\pi^2 n^2}{T^2}
  \qquad;\ n \in \mathbbm{N}\setminus\{0\}
\end{equation}
and for the case of the matrix $\gamma^2$ will be denoted by $a_j$, 
i.e.~$\gamma^2 \ket{v_j} = a_j \ket{v_j}$. With these definitions we have
\begin{equation}
  G^{\scriptscriptstyle\mathrm{(00)}} = \sum_{n,j}\;
  \frac{1}{\frac{1}{2}(c_n-a_j)} \,\ket{f_n}\ket{v_j}\bra{v_j}\bra{f_n}
\end{equation}
and thus
\begin{equation}
  G^{\scriptscriptstyle\mathrm{(00)}}(\tau,\tau^{\prime})  = 
  \bra{\tau} G^{\scriptscriptstyle\mathrm{(00)}} \ket{\tau^{\prime}}
  =
  \sum_j 
  \left(
    \sum_n \frac{2 f_n(\tau) f_n^{*}(\tau^{\prime})}{c_n - a_j}
  \right)
  \ket{v_j}\bra{v_j}
  =
  \sum_{n=1}^{\infty} \frac{2 f_n(\tau) f_n^{*}(\tau^{\prime})}
                           {c_n\one - \gamma^2}
\end{equation}
Inserting in the eigenfunctions and eigenvalues, eqs.~(\ref{eigenfuncs}) 
and (\ref{eigenvals}), and using the relation
\begin{equation} \label{eq:sumOfcos}
  \sum_{n=1}^{\infty} \frac{\cos nx}{n^2 + \alpha^2}
  =
  \frac{\pi}{2\alpha} \frac{\cosh\alpha (\pi - x)}{\sinh\alpha\pi}
  - \frac{1}{2\alpha^2}
  \qquad ;\ 0 \leq x \leq 2\pi
\end{equation}
(\cite{Gradshteyn}, eq.~1.445.2), we finally arrive at
\begin{equation}
  \label{eq:GreenFct:0-0}
  G^{\scriptscriptstyle\mathrm{(00)}}(\tau,\tau^{\prime}) = 
  \frac{ \cosh \gamma (T - |\tau+\tau^{\prime}|)}
       {\gamma\sinh \gamma T}
  -\frac{\cosh \gamma (T - |\tau-\tau^{\prime}|)}
       {\gamma\sinh \gamma T}
\end{equation}
which is defined for $0 \leq \tau,\tau^{\prime} \leq T$ and indeed satisfies
eqs.~(\ref{green1}) to (\ref{green3}) as can be verified by direct
inspection.

To calculate the remaining quadratic path integral in 
eq.~(\ref{EffAction:0-0}) we play the usual trick of multiplying and 
dividing by the known\footnote{this integral is standard in world-line
calculations and can e.g.~be evaluated by directly integrating a discretized
representation of the path integral} free path integral
\begin{equation}
\int\limits_{\tilde{y}(0)=0}^{\tilde{y}(T)=0} \!\!\!\!\mathcal{D}\tilde{y}
\;\exp\Biggl\{ -\int\limits_{0}^{T}\!\! d\tau 
\biggl[
  -\frac{1}{2}\,\tilde{y}_{\mu}(\tau)
      \!\left( \frac{1}{2}\,\delta_{\mu\nu}\partial_{\tau}^{2}  
      \right)\!   \tilde{y}_{\nu}(\tau)
\biggr]\Biggr\} = (4\pi T)^{-D/2}
\end{equation}
and calculating the resulting quotient of two functional determinants. Thus
\begin{eqnarray}
I(\gamma^2,T) & = &
\int\limits_{\tilde{y}(0)=0}^{\tilde{y}(T)=0} \!\!\!\!\mathcal{D}\tilde{y}
\;\exp\Biggl\{ -\int\limits_{0}^{T} \!\!d\tau 
\biggl[
  -\frac{1}{2}\,\tilde{y}_{\mu}(\tau)
      \!\left( \frac{1}{2}\,\delta_{\mu\nu}\partial_{\tau}^{2}  
            -\frac{1}{2}\,\gamma_{\mu\nu}^{2}
      \right)\!   \tilde{y}_{\nu}(\tau)
\biggr]
\Biggr\} \nonumber\\
& = & 
(4\pi T)^{-D/2}\; 
\frac{\textrm{Det}^{-1/2}(-\one\,\partial_{\tau}^2 +\gamma^2)}
     {\textrm{Det}^{-1/2}(-\one\,\partial_{\tau}^2)}
\end{eqnarray}
With the identity (\cite{Gradshteyn}, eq.~1.431.2)
\begin{equation} \label{eq:ProdFormulaSinh}
  \sinh x = x \cdot \prod_{n=1}^{\infty} 
  \left(
    1 + \frac{x^2}{\pi^2 n^2}
  \right)
\end{equation}
and the eigenvalues of $\partial_{\tau}^2$ given in 
eq.~(\ref{eigenvals}) we find
\begin{eqnarray} \label{eq:quadrInt:0-0}
(4\pi T)^{D/2}\; I(\gamma^2,T) & = &
\biggl[
\prod_{j=1}^D
\prod_{n=1}^{\infty}
\left(
  1 + \frac{a_j T^2}{\pi^2 n^2}
\right)
\biggr]^{-1/2}
= 
\ \biggl[
  \prod_{j=1}^D
  \frac{\sinh \sqrt{a_j T^2}}{\sqrt{a_j T^2}}
\biggr]^{-1/2}
\nonumber\\
& = & 
\!\exp\biggl\{
-\frac{1}{2} \sum_{j=1}^D \,\ln\,\frac{\sinh \sqrt{a_j T^2}}{\sqrt{a_j T^2}}
\biggr\}
\;=\;
\exp\biggl\{
-\frac{1}{2}\,\textrm{tr}\,\ln\,\frac{\sinh \gamma T}{\gamma T}
\biggr\} \qquad
\end{eqnarray}
The trace in the last expression is an ordinary matrix trace in
D-dimensional Euclidean space.

We can now join together our results for the Green's function and the
quadratic path integral, eqs.~(\ref{eq:GreenFct:0-0}) and 
(\ref{eq:quadrInt:0-0}), to complete expression (\ref{EffAction:0-0}).
With the integrated Green's function
\begin{equation}
  \int\limits_0^T\!\!d\tau d\tau^{\prime}\,
  G^{\scriptscriptstyle\mathrm{(00)}}(\tau,\tau^{\prime})
  =
  -2
  \left(
    \frac{T}{\gamma^2} - \frac{2}{\gamma^3} \tanh \frac{\gamma T}{2}
  \right)
\end{equation}
we obtain the one-loop effective action with the contribution of terms
up to second order in the derivative expansion resummed 
\begin{equation} \label{eq:EffLag:0-0}
  \Gamma^{(1)}_{\mathrm{res.}}[\varphi] = 
  \int\!d^D\!q\;\,\mathcal{L}^{(1)}_{\mathrm{res.}}[\varphi] =
  \int\!d^D\!q \int\limits_0^{\infty}\!dT\;
  \mathcal{M}^{(1)}_{\mathrm{res.}}[\varphi]
\end{equation}
where
\begin{equation} \label{eq:EffInt:0-0}
  \mathcal{M}^{(1)}_{\mathrm{res.}}[\varphi]  =  
  \frac{1}{2(4\pi)^{D/2}} \;
  \frac{e^{-\alpha T}}{T^{1+D/2}} \; 
  \exp\left\{
    \beta_{\mu}
    \left(
      \frac{T}{\gamma^2} - \frac{2}{\gamma^3}\tanh \frac{\gamma T}{2}
    \right)_{\mu\nu}
    \beta_{\nu}
    -\frac{1}{2}\,\mathrm{tr}\,\ln \frac{\sinh\gamma T}{\gamma T}
  \right\}
\end{equation}
and
\begin{displaymath}
  \alpha = m^2 + \tilde{V}(q)
  \ \qquad
  \beta_{ \mu} = \partial_{\mu}\!\left. \tilde{V}(x) \right|_{q}
  \ \qquad
  \left(\gamma^2 \right)_{\mu\nu} = 
  \ 2 \,\partial_{\mu}\partial_{\nu}\! \left. \tilde{V}(x) \right|_{q}
\end{displaymath}
with $\tilde{V}(q)=V^{\prime\prime}(\varphi(q))$
as defined in eqs.~(\ref{ak1}) and (\ref{ak2}). The complete one-loop
effective action including the higher derivative contributions can then
be written
\begin{equation} \label{eq:Full1Lp:0-0}
  \Gamma^{(1)}[\varphi] = 
  \int\!d^D\!q \int\limits_0^{\infty}\!dT\;
  \mathcal{M}^{(1)}_{\mathrm{res.}}[\varphi]
  \left\langle
    \exp\biggl\{
      -\int\limits_0^T \!d\tau\,
      \V_3\!\left(\tilde{y}(\tau)+\eta(\tau)\right)
    \biggr\}
  \right\rangle
\end{equation}
where $\V_3$ and $\eta$ have been defined in eqs.~(\ref{ak3}) and
(\ref{eta}) respectively and Wick contractions are to be performed 
by means of the resummed Green's function (\ref{eq:GreenFct:0-0})
\begin{equation}
  \left\langle
    \,\tilde{y}_{\mu}(\tau)\, \tilde{y}_{\nu}(\tau^{\prime})\,
  \right\rangle
  = - \,G^{\scriptscriptstyle\mathrm{(00)}}_{\,\mu\nu}(\tau,\tau^{\prime})
\end{equation}
Since at first glance it is not completely obvious, 
we would like to comment in the rest of this 
section on the equivalence of our result for 
the resummed one-loop effective Lagrangian, eqs.~(\ref{eq:EffLag:0-0}) and
(\ref{eq:EffInt:0-0}), to the one obtained in ref.~\cite{Masso:2001zh}. 

For this purpose, it is easiest to start with the expression 
from ref.~\cite{Masso:2001zh} (labelled $\mathcal{L}_2$ therein). 
Working in \emph{Minkowski} space-time they found with $\hbar$ set 
to unity \begin{equation} \label{eq:ResultMassoRota}
  \mathcal{L}_2 = \frac{1}{2(4\pi)^2} \int\limits_0^{\infty}\!\!ds
  \,\frac{1}{s^3}
  \left[
    \frac{e^{-\alpha s} s^2}{\sqrt{\det A_2}}
    e^{-\frac{1}{4} B_{2\mu} (A_2^{-1})^{\mu}_{\ \nu} B_2^{\nu} + C_2} 
    -e^{-m^2 s}
  \right]
\end{equation}
where
\begin{eqnarray*}
  (A_2)^{\mu}_{\ \nu} & = & (\gamma^{-1})^{\mu}_{\ \rho} 
                            (\tan\gamma s)^{\rho}_{\ \nu}\\
  B_2^{\mu} & = & -2i (\gamma^{-2})^{\mu}_{\ \rho}
                   [g^{\rho}_{\ \sigma} - (\sec\gamma s)^{\rho}_{\ \sigma}]
                   \beta^{\sigma}\\
  C_2 & = & -\frac{1}{2}\,\mathrm{tr}\ln(\cos\gamma s)
            -\beta_{\mu} (\gamma^{-3})^{\mu}_{\ \rho}
            (\tan\gamma s - \gamma s)^{\rho}_{\ \nu} \beta^{\nu}
\end{eqnarray*}
and $\alpha = m^2 + \V(q)$ as above. Any matrices without 
indices, e.g.~showing 
up as arguments of the trigonometric functions or in the determinant,
have to be understood as build up from the mixed components of the 
corresponding Lorentz tensors, i.e.~first index contravariant, second
index covariant.

Now, using $\ln\det A_2 = \mathrm{tr}\ln A_2$ to lift $A_2$ into the 
exponent and noting $s^2 = e^{\frac{1}{2}\mathrm{tr}\ln (s\mathbbm{1})}$
one can nicely combine the determinant with the $\mathrm{tr}\ln$ term in
$C_2$
\begin{equation}
  \frac{s^2}{\sqrt{\det A_2}} \,e^{C_2} =
  \exp\left\{
    -\frac{1}{2}\,\mathrm{tr}\ln\frac{\sin\gamma s}{\gamma s}
  \right\}
    e^{-\beta_{\mu} (\gamma^{-3})^{\mu}_{\ \rho}
        (\tan\gamma s - \gamma s)^{\rho}_{\ \nu} \beta^{\nu}}
\end{equation}
The remaining exponential can be written
\begin{equation}
  e^{-\frac{1}{4} B_{2\mu} (A_2^{-1})^{\mu}_{\ \nu} B_2^{\nu}}
  =
  \exp\left\{
    \beta_{\mu}\left[
      \gamma^{-3} \frac{(1-\sec\gamma s)^2}{\tan\gamma s}
    \right]^{\mu}_{\ \nu} \beta^{\nu}
  \right\}
\end{equation}
where the symmetry of $\gamma^2_{\mu\nu}$ leading to 
$f(\gamma^2)^{\mu}_{\ \nu} = f(\gamma^2)_{\nu}^{\ \mu}$ was used.
Putting everything together and noting
\begin{equation}
  \frac{(1-\sec\gamma s)^2}{\tan\gamma s} - \tan\gamma s
  =
  -2\tan\frac{\gamma s}{2}
\end{equation}
one arrives at
\begin{equation}
  \mathcal{L}_2  =  
  \frac{1}{2(4\pi)^2}
  \int\limits_0^{\infty} 
  \frac{ds}{s^3} \; 
    e^{-\alpha s}
  \exp\left\{
    \beta_{\mu}
    \left(
      \frac{s}{\gamma^2} - \frac{2}{\gamma^3}\tan \frac{\gamma s}{2}
    \right)^{\mu}_{\ \nu}
    \beta^{\nu}
    -\frac{1}{2}\,\mathrm{tr}\,\ln \frac{\sin\gamma s}{\gamma s}
  \right\}
\end{equation}
where the field independent $e^{-m^2s}$ term that was present 
in eq.~(\ref{eq:ResultMassoRota}) and acts as an infrared regulator
has been suppressed as in the world-line representation (\ref{wpi}).

For the final transition from \emph{Minkowski} to \emph{Euclidean} space
it is essential to remember that $\gamma^2$ is really the fundamental object 
in our equations, not $\gamma$. In fact, one can easily convince oneself
that all expressions above when expanded in a power series contain only
integer powers of $\gamma^2$. Writing the fundamental matrix as $\gamma^2$
rather than $\gamma$ is just a device to have such nice shorthand notations
as $\gamma^{-3} \tan(\gamma s/2)$ for a complicated power series 
in $\gamma^2$. With
\begin{math}
  \tan x = \sum_{n=0}^{\infty} \alpha_n x^{2n+1}
\end{math}
we have indeed
\begin{equation}
  \frac{2}{\gamma^3} \tan\frac{\gamma s}{2}
  =
  2 \sum_{n=0}^{\infty} \alpha_n (\gamma^{2})^{n-1} 
    \left(\frac{s}{2}\right)^{2n+1}
\end{equation}
The product of $n-1$ of the $\gamma^2$ matrices contains $n-2$ contracted
pairs of Lorentz indices, and therefore transition from \emph{Minkowski} to 
\emph{Euclidean} space-time gives $n-2$ minus signs:
\begin{equation}
  2 \sum_{n=0}^{\infty} \alpha_n 
  (-1)^{n-2}(\gamma^{2})^{n-1} \left(\frac{s}{2}\right)^{2n+1}
  =
  \frac{2}{\gamma^3} \sum_{n=0}^{\infty} \alpha_n (-1)^n 
  \left(\frac{\gamma s}{2}\right)^{2n+1}
  =
  \frac{2}{\gamma^3} \tanh\frac{\gamma s}{2}
\end{equation}
Hence, the fact that $\gamma^2$ is the fundamental object and not $\gamma$ 
guarantees that we only get a sign change in every second power
of $\gamma$ and thus a transition from \emph{tan} to 
\emph{tanh}.\footnote{It is amusing that the same effect is usually 
triggered by an $i$ coming into play by Wick rotating the Schwinger
proper time.} In the same way the \emph{sin} translates to \emph{sinh}
with an additional minus sign compensated by the trace and we have 
thereby shown the equivalence of our result (\ref{eq:EffInt:0-0})
with eq.~(\ref{eq:ResultMassoRota}).

\section{Taking different paths} \label{second_WL_approach}
\setcounter{equation}{0}

The calculation of the previous section was based on the path integral 
representation of the one-loop effective action given in eq.~(\ref{wpi}).
This representation contains a path integral over all paths 
running from $x(0)=q$ to $x(T)=q$ followed by an ordinary integral
over all possible positions of the common starting and end point $q$.
Together, this corresponds to an integration over the full set of closed
paths in configuration space. 

We will show now that 
by choosing a different parametrization of this set a very effective way 
of resumming terms up to quadratic order in the Taylor expansion of
the potential opens up.

To this end, let us define the ``center of mass'' of a given 
path $x(\tau)$ as
\begin{equation}
  q = \frac{1}{T} \int\limits_0^T\!\!d\tau\,x(\tau)
\end{equation}
Doing so, we can alternatively cover the total set of all closed paths by 
first integrating the paths with a fixed ``center of mass'' $q$ 
and afterwards integrating all possible positions of this center.
Hence, the representation of the one-loop effective action 
(\ref{wpi}) may also be written
\begin{equation} \label{eq:COM-EffAct}
\Gamma^{(1)} [\varphi] 
 =  \frac{1}{2} \int\limits_{0}^{\infty} \frac{dT}{T}\,e^{-m^{2}T} 
    \int\!d^D\!q\;\;
    \mathcal{N}^{\,\prime} \!\!\!\!\!
    \oint\limits_{(\mathrm{CM}=q)} \!\!\!\!\mathcal{D}x\,
    \exp\Biggl\{ -\int\limits_{0}^{T}\!\! d\tau 
    \left[ \frac{1}{4}\,\dot{x}^{2}(\tau)  +
    V^{\prime\prime}
    \!\left( \varphi(x) \right) \right] \Biggr\}  
\end{equation}
The factor $\mathcal{N}^{\,\prime}$ takes care of a change in 
the measure accompanying the transition from the path integral with 
fixed end points to the path integral with fixed ``center of mass''
(see appendix \ref{app:EndPointsVersusCenter} for details).
However, since $\mathcal{N}^{\,\prime}$ will ultimately drop out of 
the calculation there is no need to specify it any further here.

Inserting again the expansion (\ref{taylor}) of the potential,
performing a shift in the integration path
\begin{math}
  y(\tau) =  x(\tau) - q 
\end{math}
where now $q$ is the ``center of mass'' and finally integrating
partially in the exponent neglecting the boundary terms leads to
\begin{eqnarray} \label{eq:PathIntRepofGamma:CM}
\lefteqn{
\Gamma^{(1)}[\varphi]  =  
\frac{1}{2} \int\limits_{0}^{\infty} \frac{dT}{T}\,e^{-m^{2}T}
\!\!\int\! d^{D}q \;e^{-{\V(q)T}}} \\
& & \times\ \;
\mathcal{N}^{\,\prime} \!\!\!\!\!
\oint\limits_{(\mathrm{CM}=0)} \!\!\!\!\mathcal{D}y\;
\exp\Biggl\{ -\int\limits_{0}^{T}\!\! d\tau 
\biggl[
  -\frac{1}{2}\,y_{\mu}(\tau)
      \left( \frac{1}{2}\,\delta_{\mu\nu}\partial_{\tau}^{2}  
            -\frac{1}{2}\,\gamma_{\mu\nu}^{2}
      \right)   y_{\nu}(\tau) 
  + \,\beta_{\mu} y_{\mu}(\tau) + \V_3(y(\tau))
\biggr]
\Biggr\} \nonumber
\end{eqnarray}
analogously to eq.~(\ref{eq:GammaFirstShift}).
However, because this time the path integral is performed over the set 
of closed paths $y(\tau)$ with a vanishing ``center of mass'',
the linear term in the exponent simply disappears being 
proportional to the integral of $y(\tau)$.

Consequently, following the present approach in resumming terms 
up to second order we not even have to calculate the Green's function.
All that remains is to determine the quadratic path integral
\begin{eqnarray} 
\tilde{I}(\gamma^2,T) & = &
\mathcal{N}^{\,\prime} \!\!\!\!\!
\oint\limits_{(\mathrm{CM}=0)} \!\!\!\!\mathcal{D}y\;
\exp\Biggl\{ -\int\limits_{0}^{T}\!\! d\tau 
\biggl[
  -\frac{1}{2}\,y_{\mu}(\tau)
      \left( \frac{1}{2}\,\delta_{\mu\nu}\partial_{\tau}^{2}  
            -\frac{1}{2}\,\gamma_{\mu\nu}^{2}
      \right)   y_{\nu}(\tau) 
\biggr]
\Biggr\} \nonumber\\ \label{eq:tildeI}
& = & 
\mathcal{N}^{\,\prime}\!\!\!\!\!
\oint\limits_{(\mathrm{CM}=0)} \!\!\!\!\mathcal{D}y\;
\exp\biggl\{
  -\int\limits_0^T \!d\tau\,
  \frac{1}{4}\,\dot{y}^2
\biggr\}\;
\frac{\textrm{Det}^{-1/2}_{\scriptscriptstyle\mathrm{\,CM=0}}
      (-\one\,\partial_{\tau}^2 +\gamma^2)}
     {\textrm{Det}^{-1/2}_{\scriptscriptstyle\mathrm{\,CM=0}}
      (-\one\,\partial_{\tau}^2)}
\end{eqnarray}
At this point the factor $\mathcal{N}^{\,\prime}$ is reabsorbed by the free
path integral transforming back to the free integral with fixed
end points which is known to evaluate to $(4\pi T)^{-D/2}$
(see appendix \ref{app:EndPointsVersusCenter}).

The functional determinants are defined on the subspace of
closed paths with vanishing ``center of mass''.
A complete set of real valued and normalized one-dimensional
eigenfunctions of $\partial_{\tau}^2$ with ``center of mass''
zero and with equal values at $\tau=0$ and $\tau=T$ is given by
\begin{equation} \label{eq:eigenstatesCM}
\left.
\begin{array}{rcccl}
  \bigl\langle\hspace{0.5pt}\tau\hspace{0.5pt}|
              \hspace{0.5pt}\tilde{f}_n\hspace{0.5pt}\bigr\rangle &=& 
  \tilde{f}_n(\tau) & = & \frac{-i}{\sqrt{2T}}
              \left(
                e^{+2\pi i n \tau/T} - e^{-2\pi i n \tau/T}
              \right) \\[2.0ex]
  \bigl\langle\hspace{0.5pt}\tau\hspace{0.5pt}|
              \hspace{0.5pt}\tilde{g}_n\hspace{0.5pt}\bigr\rangle &=& 
  \tilde{g}_n(\tau) & = & \frac{1}{\sqrt{2T}}
              \left(
                e^{+2\pi i n \tau/T} + e^{-2\pi i n \tau/T}
              \right)
\end{array}
\right\} \qquad;\ n \in \mathbbm{N}\setminus\{0\}
\end{equation}
Hence, there are two linear independent eigenfunctions to any 
eigenvalue, i.e.
\begin{math}
  \partial_{\tau}^2 \tilde{f}_n(\tau) = \tilde{c}_n \tilde{f}_n(\tau)
\end{math}
and
\begin{math}
  \partial_{\tau}^2 \tilde{g}_n(\tau) = \tilde{c}_n \tilde{g}_n(\tau)
\end{math}
with
\begin{equation} \label{eigenvalsCM}
  \tilde{c}_n = - \frac{4\pi^2 n^2}{T^2}
  \qquad;\ n \in \mathbbm{N}\setminus\{0\}
\end{equation}

Using in addition the normalized eigenvectors 
$\ket{v_1},\dots,\ket{v_D}$ 
of $\gamma^2$ with $\gamma^2 \ket{v_j} = a_j \ket{v_j}$ that 
we already introduced in section \ref{first_WL_approach}, the 
quotient of the two determinants occurring in eq.~(\ref{eq:tildeI}) can
be evaluated. One finds
\begin{eqnarray}
(4\pi T)^{D/2}\; \tilde{I}(\gamma^2,T) & = &
\biggl[
\prod_{j=1}^D
\prod_{n=1}^{\infty}
\left(
  1 + \frac{a_j T^2}{4\pi^2 n^2}
\right)^{\!\!2}\,
\biggr]^{-1/2}
= 
\ \biggl[
  \prod_{j=1}^D
  \frac{\sinh (\sqrt{a_j T^2}/2)}{\sqrt{a_j T^2}/2}
\biggr]^{-1}
\nonumber\\
& = & 
\!\exp\biggl\{
-\sum_{j=1}^D \,\ln\,\frac{\sinh (\sqrt{a_j T^2}/2)}{\sqrt{a_j T^2}/2}
\biggr\}
\;=\;
\exp\biggl\{
-\,\textrm{tr}\,\ln\,\frac{\sinh (\gamma T/2)}{\gamma T/2}
\biggr\} \qquad
\end{eqnarray}
where we again have taken advantage of the identity 
(\ref{eq:ProdFormulaSinh}).

Thus, for the one-loop effective action with terms up to second order 
in the derivatives resummed we obtain the alternative expression 
\begin{equation} \label{eq:EffLag:CM=0}
  \tilde{\Gamma}^{(1)}_{\mathrm{res.}}[\varphi] = 
  \int\!d^D\!q\;\,\tilde{\mathcal{L}}^{(1)}_{\mathrm{res.}}[\varphi] =
  \int\!d^D\!q \int\limits_0^{\infty}\!dT\;
  \tilde{\mathcal{M}}^{(1)}_{\mathrm{res.}}[\varphi]
\end{equation}
with
\begin{equation} \label{eq:EffInt:CM=0}
  \tilde{\mathcal{M}}^{(1)}_{\mathrm{res.}}[\varphi]  =  
  \frac{1}{2(4\pi)^{D/2}} \;
  \frac{e^{-\alpha T}}{T^{1+D/2}} \; 
  \exp\left\{
    -\,\mathrm{tr}\,\ln \frac{\sinh(\gamma T/2)}{\gamma T/2}
  \right\}
\end{equation}
and the same abbreviations as used in eq.~(\ref{eq:EffInt:0-0}).
Accordingly, the complete one-loop effective action may alternatively 
be written
\begin{equation} \label{eq:Full1Lp:CM}
  \Gamma^{(1)}[\varphi] = 
  \int\!d^D\!q \int\limits_0^{\infty}\!dT\;
  \tilde{\mathcal{M}}^{(1)}_{\mathrm{res.}}[\varphi]
  \left\langle
    \exp\biggl\{
      -\int\limits_0^T \!d\tau\,
      \V_3\!\left(y(\tau)\right)
    \biggr\}
  \right\rangle
\end{equation}
where now contractions of the world-line fields are to be performed
with the appropriate Green's function
\begin{equation} \label{eq:ContractionCM}
  \left\langle
    \,y_{\mu}(\tau)\, y_{\nu}(\tau^{\prime})\,
  \right\rangle
  = - \,G^{\scriptscriptstyle\mathrm{(CM)}}_{\;\mu\nu}(\tau,\tau^{\prime})
\end{equation}
inverting still the operator  
\begin{math}
  K = \frac{1}{2}\left(\one\,\partial_{\tau}^{2} - \gamma^{2}\right)
\end{math}
but this time on the space of closed paths with ``center of mass'' zero.
The construction of $G^{\scriptscriptstyle\mathrm{(CM)}}$
can be carried out along the same lines as 
$G^{\scriptscriptstyle\mathrm{(00)}}$ was obtained in the previous 
section. 

In terms of the eigenstates $\bigl|\tilde{f}_n\bigr\rangle$
and $\bigl|\tilde{g}_n\bigr\rangle$ of the operator $\partial_{\tau}^2$ 
in the subspace of functions with ``center of mass'' zero, 
eq.~(\ref{eq:eigenstatesCM}), and the eigenstates $\ket{v_j}$ of 
the matrix $\gamma^2$ one has
\begin{equation}
  G^{\scriptscriptstyle\mathrm{(CM)}} = \sum_{j=1}^D\sum_{n=1}^{\infty}\;
  \frac{1}{\frac{1}{2}(\tilde{c}_n-a_j)}
  \left[\,
  \bigl|\tilde{f}_n\bigr\rangle
  \bigl|v_j\bigr\rangle
  \bigl\langle v_j\bigr|
  \bigl\langle\tilde{f}_n\bigr|
  \ +\ 
  \bigl|\tilde{g}_n\bigr\rangle
  \bigl|v_j\bigr\rangle
  \bigl\langle v_j\bigr|
  \bigl\langle\tilde{g}_n\bigr|
  \,\right]
\end{equation}
and thus
\begin{equation}
  G^{\scriptscriptstyle\mathrm{(CM)}}(\tau,\tau^{\prime})  = 
  \bra{\tau} G^{\scriptscriptstyle\mathrm{(CM)}} \ket{\tau^{\prime}}
  =
  \sum_{n=1}^{\infty} \frac{2\, 
    [\tilde{f}_n(\tau) \tilde{f}_n^{*}(\tau^{\prime}) \;+\;
     \tilde{g}_n(\tau) \tilde{g}_n^{*}(\tau^{\prime})   ]}
  {\tilde{c}_n\one - \gamma^2}
\end{equation}
Finally, inserting the functions $\tilde{f}_n(\tau)$ and $\tilde{g}_n(\tau)$
from eq.~(\ref{eq:eigenstatesCM}) leading to
\begin{equation}
  \tilde{f}_n(\tau) \tilde{f}_n^{*}(\tau^{\prime}) \;+\;
  \tilde{g}_n(\tau) \tilde{g}_n^{*}(\tau^{\prime})
  =
  \frac{2}{T}\,\cos 
  \Bigl(2\pi n \,\frac{\tau-\tau^{\prime}}{T}\Bigr)
\end{equation}
and using for a second time the relation (\ref{eq:sumOfcos}) and
the eigenvalues $\tilde{c}_n$ from eq.~(\ref{eigenvalsCM}) yields
\begin{equation} \label{eq:GreenFct:CM}
  G^{\scriptscriptstyle\mathrm{(CM)}}(\tau,\tau^{\prime})  =
  -\,\frac{\cosh \gamma (T/2 - |\tau-\tau^{\prime}|)}
          {\gamma\sinh\,(\gamma T/2)}
  \ +\  \frac{2}{\gamma^2 T}
\end{equation}
Note that 
\begin{equation}
  \frac{1}{2}\left(\one\,\partial_{\tau}^{2} - \gamma^{2}\right)
  G^{\scriptscriptstyle\mathrm{(CM)}}(\tau,\tau^{\prime})
  = \one \Bigl( \delta(\tau-\tau^{\prime}) -\frac{1}{T}\Bigr)
\end{equation}
Hence, $G^{\scriptscriptstyle\mathrm{(CM)}}(\tau,\tau^{\prime})$ 
as given in eq.~(\ref{eq:GreenFct:CM}) indeed only inverts the 
operator $K$ in the subspace of paths with a vanishing 
``center of mass''. However, this is all we need. 
In combination with
\begin{math}
  G^{\scriptscriptstyle\mathrm{(CM)}}_{\;\mu\nu}(\tau,\tau^{\prime}) =
  G^{\scriptscriptstyle\mathrm{(CM)}}_{\;\nu\mu}(\tau^{\prime},\tau)
\end{math}\,,
\begin{math}
  G^{\scriptscriptstyle\mathrm{(CM)}}_{\;\mu\nu}(0,\tau^{\prime}) =
  G^{\scriptscriptstyle\mathrm{(CM)}}_{\;\mu\nu}(T,\tau^{\prime})
\end{math}\,,
\begin{math}
  \dot{G}^{\scriptscriptstyle\mathrm{(CM)}}_{\;\mu\nu}(0,\tau^{\prime}) =
  \dot{G}^{\scriptscriptstyle\mathrm{(CM)}}_{\;\mu\nu}(T,\tau^{\prime})
\end{math}
(dots indicating derivatives with respect to the first argument) and
\begin{math}
  \int_0^T\!\!d\tau\,
  G^{\scriptscriptstyle\mathrm{(CM)}}_{\;\mu\nu}(\tau,\tau^{\prime}) =0
\end{math},
it has all the properties necessary to justify 
eq.~(\ref{eq:ContractionCM}).

\section{Comparison with the ``inverse mass expansion''}
\label{inverseMass}
\setcounter{equation}{0}

In the previous two sections we derived two expressions for the one-loop
effective action, eqs.~(\ref{eq:Full1Lp:0-0}) and (\ref{eq:Full1Lp:CM}).
Each of them consists of a resummed part containing the total contribution 
of a certain subclass of terms and a remainder of individual terms still
to be summed up by hand in the form of Wick contractions of the world-line
fields.

In order to clarify the relation between the two resummed effective actions
$\Gamma^{(1)}_{\mathrm{res.}}$ and $\tilde{\Gamma}^{(1)}_{\mathrm{res.}}$,
eqs.~(\ref{eq:EffLag:0-0}), (\ref{eq:EffInt:0-0}) and (\ref{eq:EffLag:CM=0}), 
(\ref{eq:EffInt:CM=0}) respectively, and to have another crosscheck of 
the calculation we now compare our results to the ``inverse mass expansion'' 
discussed in refs.~\cite{Fliegner:1994zc,Fliegner:1997rk}.

The ``inverse mass expansion'' is an expansion of the effective action 
in powers of the parameter $T$ which has mass dimension $-2$. For the 
case of a real scalar field that we are considering here, the general
results in refs.~\cite{Fliegner:1994zc,Fliegner:1997rk} reduce to
\begin{equation} \label{eq:InvMassFliegner}
  \Gamma^{(1)}[\varphi] \;=\;
  \frac{1}{2}
  \int\limits_0^{\infty}\frac{dT}{T}
  \frac{e^{-m^2T}}{(4\pi T)^{D/2}}
  \sum_{n=0}^{\infty} \frac{(-T)^n}{n!}
  \int\!\!d^D\!q\;\hat{O}_n
\end{equation}
where the $\hat{O}_n$ up to $n=5$ are given by
\begin{equation} \label{eq:InvMassCoeffFliegner}
\begin{array}{lcl}
  \hat{O}_0 &=& 1\\[0.75ex]
  \hat{O}_1 &=& \V\\[0.75ex]
  \hat{O}_2 &=& \V^2\\[0.75ex]
  \hat{O}_3 &=& \V^3 + \frac{1}{2}\V_{\mu}\V_{\mu}\\[0.75ex]
  \hat{O}_4 &=& \V^4 + 2 \V \V_{\mu}\V_{\mu} 
                + \frac{1}{5}\V_{\mu\nu}\V_{\mu\nu}\\[0.75ex] 
  \hat{O}_5 &=& \V^5 + 5 \V^2\V_{\mu}\V_{\mu} + \V\V_{\mu\nu}\V_{\mu\nu}
                + \frac{5}{3} \V_{\mu}\V_{\nu}\V_{\mu\nu}
                + \frac{1}{14} \V_{\kappa\mu\nu}\V_{\kappa\mu\nu}
\end{array}
\end{equation}
and the shorthand notation $\V_{\mu}=\partial_{\mu}\V$, 
$\V_{\mu\nu}=\partial_{\mu}\partial_{\nu}\V$ etc.~is used
with $\V$ and its derivatives evaluated at the point $q$.

In comparing $\Gamma^{(1)}_{\mathrm{res.}}$ and 
$\tilde{\Gamma}^{(1)}_{\mathrm{res.}}$ with the ``invers mass expansion''
one has to take into account that (\ref{eq:InvMassFliegner}) is an 
expansion of the \emph{full} one-loop effective action, whereas the 
resummed effective actions $\Gamma^{(1)}_{\mathrm{res.}}$ and 
$\tilde{\Gamma}^{(1)}_{\mathrm{res.}}$ only contain certain subclasses
of terms. Hence, an expansion of $\Gamma^{(1)}_{\mathrm{res.}}$ or
$\tilde{\Gamma}^{(1)}_{\mathrm{res.}}$ in powers of $T$ will in general 
not lead to the coefficients (\ref{eq:InvMassCoeffFliegner}) belonging to the
full one-loop effective action.

However, if we confine ourselves to the case where third and higher 
derivatives of $\V$ vanish, then the term $\V_3$ 
in eqs.~(\ref{eq:Full1Lp:0-0}) and (\ref{eq:Full1Lp:CM}) is zero 
and both resummed expressions $\Gamma^{(1)}_{\mathrm{res.}}$ and 
$\tilde{\Gamma}^{(1)}_{\mathrm{res.}}$ coincide with the full one-loop
effective action in this case. 
Consequently, the expansions of $\Gamma^{(1)}_{\mathrm{res.}}$ and  
$\tilde{\Gamma}^{(1)}_{\mathrm{res.}}$ in powers of $T$ should lead 
to coefficients, at least consistent with (\ref{eq:InvMassCoeffFliegner}) 
under the assumption of vanishing third and higher derivatives of $\V$.

Let us check, if this is true. Expanding $\Gamma^{(1)}_{\mathrm{res.}}$
analogously to the expansion of $\Gamma^{(1)}$ in 
eq.~(\ref{eq:InvMassFliegner}) yields
\begin{equation} \label{eq:InvMassCoeff0-0}
\begin{array}{lcl}
  \hspace{-1.0ex}O_0 &=& 1\\[0.75ex]
  \hspace{-1.0ex}O_1 &=& \V\\[0.75ex]
  \hspace{-1.0ex}O_2 &=& \V^2 -\frac{1}{3}\V_{\mu\mu}\\[0.75ex]
  \hspace{-1.0ex}O_3 &=& \V^3 -\V\V_{\mu\mu} - 
                       \frac{1}{2}\V_{\mu}\V_{\mu}\\[0.75ex]
  \hspace{-1.0ex}O_4 &=& \V^4 -2\V^2\V_{\mu\mu} -2\V\V_{\mu}\V_{\mu} 
          + \frac{1}{3}\V_{\mu\mu}\V_{\nu\nu}
          + \frac{4}{15}\V_{\mu\nu}\V_{\mu\nu}
          \\[0.75ex] 
  \hspace{-1.0ex}O_5 &=& \V^5 - \frac{10}{3}\V^3\V_{\mu\mu}
          - 5 \V^2\V_{\mu}\V_{\mu} 
          + \frac{5}{3}\V\V_{\mu\mu}\V_{\nu\nu}
          + \frac{4}{3}\V\V_{\mu\nu}\V_{\mu\nu}
          + \frac{5}{3}\V_{\mu}\V_{\mu}\V_{\nu\nu}
          + 2\V_{\mu}\V_{\nu}\V_{\mu\nu}
\end{array}
\end{equation}
Accordingly, $\tilde{\Gamma}^{(1)}_{\mathrm{res.}}$ leads to the
coefficients 
\begin{equation} \label{eq:InvMassCoeffCM}
\begin{array}{lcl}
  \hspace{-1.0ex}\tilde{O}_0 &=& 1\\[0.75ex]
  \hspace{-1.0ex}\tilde{O}_1 &=& \V\\[0.75ex]
  \hspace{-1.0ex}\tilde{O}_2 &=& \V^2 -\frac{1}{6}\V_{\mu\mu}\\[0.75ex]
  \hspace{-1.0ex}\tilde{O}_3 &=& \V^3 -\frac{1}{2}\V\V_{\mu\mu}\\[0.75ex]
  \hspace{-1.0ex}\tilde{O}_4 &=& \V^4 -\V^2\V_{\mu\mu}
                  + \frac{1}{12}\V_{\mu\mu}\V_{\nu\nu}
                  + \frac{1}{30}\V_{\mu\nu}\V_{\mu\nu}\\[0.75ex] 
  \hspace{-1.0ex}\tilde{O}_5 &=& \V^5 - \frac{5}{3}\V^3\V_{\mu\mu}
                  + \frac{5}{12}\V\V_{\mu\mu}\V_{\nu\nu}
                  + \frac{1}{6}\V\V_{\mu\nu}\V_{\mu\nu}
  \hspace{38ex}
\end{array}
\end{equation}
To compare eqs.~(\ref{eq:InvMassCoeffFliegner})--(\ref{eq:InvMassCoeffCM})
one has to recall that these coefficients appear under an integral over
$q$ in eq.~(\ref{eq:InvMassFliegner}) which means that apparently different 
looking terms can be transformed into each other by partial integrations.

The $\V_{\mu\mu}$ term in $O_2$ and $\tilde{O}_2$, for instance, is a 
boundary term with respect to this integration. Therefore we have in fact
$\hat{O}_2 \cong O_2 \cong \tilde{O}_2$, denoting by the symbol `\,$\cong$\,'
equality up to partial integrations.
 In the same way we find
\begin{equation}
  -\V\V_{\mu\mu} -{\textstyle\frac{1}{2}}\V_{\mu}\V_{\mu}
  \quad\cong\quad
  -\V\V_{\mu\mu}
  +{\textstyle\frac{1}{2}}\V\V_{\mu\mu}
  = -{\textstyle\frac{1}{2}}\V\V_{\mu\mu}
  \quad\cong\quad
  +\,{\textstyle\frac{1}{2}}\V_{\mu}\V_{\mu}
\end{equation}
i.e.~$O_3 \cong \tilde{O}_3 \cong \hat{O}_3$.
Inspecting the fourth order terms we can use 
$\V^2\V_{\mu\mu}\cong -2\V\V_{\mu}\V_{\mu}$ and the fact that
under the assumption of vanishing third derivatives 
$\V_{\mu\nu}\V_{\mu\nu} \cong 0$ as well 
as $\V_{\mu\mu}\V_{\nu\nu} \cong 0$
to establish $O_4 \cong \tilde{O}_4 \cong \hat{O}_4$.
Finally, using $\V^3\V_{\mu\mu}\cong -3\V^2V_{\mu}V_{\mu}$ and the three
replacements (valid for third derivatives set to zero) 
$\V\V_{\mu\nu}\V_{\mu\nu} \cong -V_{\mu}V_{\nu}\V_{\mu\nu}$,
$\V\V_{\mu\mu}\V_{\nu\nu} \cong +2V_{\mu}V_{\nu}\V_{\mu\nu}$ and
$\V_{\mu}\V_{\mu}\V_{\nu\nu} \cong -2V_{\mu}V_{\nu}\V_{\mu\nu}$,
one finds $O_5 \cong \tilde{O}_5 \cong \hat{O}_5$.

Thus, we have shown that under the assumption of vanishing third and higher
derivatives the first non-trivial terms of the expansions
of $\Gamma^{(1)}_{\mathrm{res.}}$ and 
$\tilde{\Gamma}^{(1)}_{\mathrm{res.}}$ in powers of $T$ do indeed
coincide with the ``inverse mass expansion'' (\ref{eq:InvMassFliegner}),
(\ref{eq:InvMassCoeffFliegner}) of the full one-loop effective action.
This is in agreement with the relation
\begin{math}
  \Gamma^{(1)} =
  \Gamma^{(1)}_{\mathrm{res.}} =
  \tilde{\Gamma}^{(1)}_{\mathrm{res.}}
\end{math}
being a simple consequence of eqs.~(\ref{eq:Full1Lp:0-0}) 
and (\ref{eq:Full1Lp:CM}) in this case
and it serves as a crosscheck that we did not make any 
calculational errors in deriving the explicit expressions for the 
resummed effective actions, eqs.~(\ref{eq:EffLag:0-0}), 
(\ref{eq:EffInt:0-0}) and (\ref{eq:EffLag:CM=0}), (\ref{eq:EffInt:CM=0}).

However, there is something else we can learn from the expansions 
(\ref{eq:InvMassCoeff0-0}) and (\ref{eq:InvMassCoeffCM}).
The assumption of vanishing third and higher derivatives was imposed 
above to open up an easy way of comparing the resummed actions
$\Gamma^{(1)}_{\mathrm{res.}}$ and $\tilde{\Gamma}^{(1)}_{\mathrm{res.}}$
to the well-known ``inverse mass expansion'' of the full one-loop 
effective action. Without this assumption we would have been forced to take
into account the contributions from the contractions of world-line
fields in eqs.~(\ref{eq:Full1Lp:0-0}) and (\ref{eq:Full1Lp:CM}).
However, in comparing $\Gamma^{(1)}_{\mathrm{res.}}$ 
and $\tilde{\Gamma}^{(1)}_{\mathrm{res.}}$ with each other, there is no 
need to confine oneself to this special case.

Therefore, let us compare the coefficients (\ref{eq:InvMassCoeff0-0}) and 
(\ref{eq:InvMassCoeffCM}) without assuming third derivatives being zero.
We have only referred to this assumption in the discussion of the 
fourth and fifth order terms, i.e.~we still have $O_n = \tilde{O}_n$ for
$n=0,1,2,3$. Using 
\begin{math}
  \V_{\mu\mu}\V_{\nu\nu} \cong
  -\V_{\mu}\V_{\mu\nu\nu} \cong
  \V_{\mu\nu}\V_{\mu\nu}
\end{math}
together with the replacements
\begin{math}
  \V_{\mu}\V_{\mu}\V_{\nu\nu} \cong -2\V_{\mu}\V_{\nu}\V_{\mu\nu}
\end{math}
and
\begin{math}
  \V\V_{\mu\nu}\V_{\mu\nu} \cong
  -\V_{\mu}\V_{\nu}\V_{\mu\nu} - \V\V_{\nu}\V_{\mu\mu\nu} \cong
  -\V_{\mu}\V_{\nu}\V_{\mu\nu} + \V_{\nu}\V_{\nu}\V_{\mu\mu} 
  +\V\V_{\nu\nu}\V_{\mu\mu} \cong
  -3\V_{\mu}\V_{\nu}\V_{\mu\nu} + \V\V_{\mu\mu}\V_{\nu\nu}
\end{math}
leads to the following two expressions for the fourth order terms
\begin{eqnarray}
  O_4 &\cong& \V^4 + 2\V\V_{\mu}\V_{\mu} + \;
              {\textstyle\frac{3}{5}}\V_{\mu\mu}\V_{\nu\nu} \\
  \tilde{O}_4 &\cong& \V^4 + 2\V\V_{\mu}\V_{\mu} +
              {\textstyle\frac{7}{60}}\V_{\mu\mu}\V_{\nu\nu}
\end{eqnarray}
and to
\begin{eqnarray}
  O_5 &\cong& \V^5 + 5\V^2\V_{\mu}\V_{\mu} + \;\,3\V\V_{\mu\mu}\V_{\nu\nu}
              -{\textstyle\frac{16}{3}}\V_{\mu}\V_{\nu}\V_{\mu\nu} \\
  \tilde{O}_5 &\cong& \V^5 + 5\V^2\V_{\mu}\V_{\mu} +
              {\textstyle\frac{7}{12}}\V\V_{\mu\mu}\V_{\nu\nu} -
              \,{\textstyle\frac{1}{2}}\V_{\mu}\V_{\nu}\V_{\mu\nu}
\end{eqnarray}
at fifth order. Thus, in general, we have 
\begin{math}
  \Gamma^{(1)}_{\mathrm{res.}}[\varphi] \neq
  \tilde{\Gamma}^{(1)}_{\mathrm{res.}}[\varphi]
\end{math}
which indicates that by the transition to the world-line integral of 
paths with ``center of mass'' zero also the class of terms
is changed that is taken into account by the resummation.

One may wonder if this does not contradict our derivation of
$\Gamma^{(1)}_{\mathrm{res.}}$ and 
$\tilde{\Gamma}^{(1)}_{\mathrm{res.}}$ which seems to suggest that 
not only the two representations of the full one-loop effective action,
eqs.~(\ref{EffAction:0-0}) and (\ref{eq:PathIntRepofGamma:CM}), should 
be equivalent, but also the two expressions for the resummed parts
which follow from the full expressions (\ref{EffAction:0-0}) and 
(\ref{eq:PathIntRepofGamma:CM}) by neglecting $\V_3$ in each 
of them.

However, there is a subtle point to observe. Indeed, starting from 
eq.~(\ref{EffAction:0-0}) without the $\V_3$ term and tracing back
the steps of the derivation one arives at 
\begin{eqnarray} \label{eq:ResGammaBack:0-0}
\lefteqn{
\Gamma^{(1)}_{\mathrm{res.}}[\varphi]  =  
\frac{1}{2} \int\limits_{0}^{\infty} \frac{dT}{T}\,e^{-m^{2}T}
\!\!\int d^D\!q \;} \\
& & \times\;
\int\limits_{x(0)=q}^{x(T)=q} \!\!\!\!\mathcal{D}x\;
\exp\Biggl\{ -\int\limits_{0}^{T}\!\! d\tau 
\biggl[
  \frac{1}{4}\,\dot{x}^2 +
  \V(q) + \beta_{\mu} \left(x  - q \right)_{\mu} 
        + \frac{1}{4}\,\gamma_{\mu\nu}^{2} 
                       \left( x - q \right)_{\mu} 
                       \left( x - q \right)_{\nu} 
\biggr]
\Biggr\} \hspace{2ex}\nonumber
\end{eqnarray}
In the same way, starting with eq.~(\ref{eq:PathIntRepofGamma:CM})
and dropping $\V_3$ leads back to
\begin{eqnarray} \label{eq:ResGammaBack:CM}
\lefteqn{
\tilde{\Gamma}^{(1)}_{\mathrm{res.}}[\varphi]  =  
\frac{1}{2} \int\limits_{0}^{\infty} \frac{dT}{T}\,e^{-m^{2}T}
\!\!\int d^D\!q \;} \\
& & \times\ \;\mathcal{N}^{\,\prime} \!\!\!\!\!
\oint\limits_{(\mathrm{CM}=q)} \!\!\!\!\mathcal{D}x\;
\exp\Biggl\{ -\int\limits_{0}^{T}\!\! d\tau 
\biggl[
  \frac{1}{4}\,\dot{x}^2 +
  \V(q) + \beta_{\mu} \left(x  - q \right)_{\mu} 
        + \frac{1}{4}\,\gamma_{\mu\nu}^{2} 
                       \left( x - q \right)_{\mu} 
                       \left( x - q \right)_{\nu} 
\biggr]
\Biggr\} \nonumber
\end{eqnarray}
Hence, changing the integration
\begin{equation} \label{eq:transition}
  \int\!\!d^D\!q\!\!\!
  \int\limits_{x(0)=q}^{x(T)=q} \!\!\!\!\mathcal{D}x
  \quad\longrightarrow\quad
  \mathcal{N}^{\,\prime}\!
  \int\!\!d^D\!q\!\!\!
  \oint\limits_{(\mathrm{CM}=q)} \!\!\!\!\mathcal{D}x
\end{equation}
would imply
\begin{math}
  \Gamma^{(1)}_{\mathrm{res.}}[\varphi] =
  \tilde{\Gamma}^{(1)}_{\mathrm{res.}}[\varphi]
\end{math} in contradiction to what we found above.

The solution to this puzzle is that the change of integration paths
(\ref{eq:transition}) is not allowed in the transition from 
eq.~(\ref{eq:ResGammaBack:0-0}) to eq.~(\ref{eq:ResGammaBack:CM}),
because the integrand has an explicit $q$ dependence.
This becomes clear from our investigations in appendix 
\ref{app:EndPointsVersusCenter} where we justify this change of 
integration paths: By assuming an additional explicit $q$ dependence 
(denoted by the last argument after the semicolon) 
eq.~(\ref{eq:DiscrQ2Q}) changes to
\begin{equation}
  \int\!\!d^D\!q\!\!\!
  \int\limits_{x(0)=q}^{x(T)=q} \!\!\!\!\mathcal{D}x\;F[x] 
  = \mathcal{N}
  \int\!\!d^D\!q
  \int\biggl[\prod_{m=1}^{N} d^D\!x_m\biggr]
  f(q,x_1,\dots,x_N,q;q) 
\end{equation}
Then, following the same steps as described in appendix 
\ref{app:EndPointsVersusCenter} one obtains
instead of eq.~(\ref{eq:StartFromFixedEndpoints})
\begin{equation}
  \int\!\!d^D\!q\!\!\!
  \int\limits_{x(0)=q}^{x(T)=q} \!\!\!\!\mathcal{D}x\;F[x] 
  = 
  \int\!\!d^D\!c\;
  \mathcal{N}^{\,\prime}\,\mathcal{N}\!\!
  \int\biggl[\prod_{m=0}^{N-1} d^D\!x_m\biggr]
  f(x_0,x_1,\dots,x_N(c,x_0,\dots,x_{N-1}),x_0;x_0)  
\end{equation}
Thus, though we have performed the transition to integration paths 
with ``center of mass'' $c$ the explicit parameter dependence of the 
integrand is still evaluated at $x_0$, not at the center $c$. Therefore
eqs.~(\ref{eq:ResGammaBack:0-0}) and (\ref{eq:ResGammaBack:CM}) are 
not equivalent in general.

On the other hand, in the case of the full one-loop effective action
though we write the potential as a Taylor expansion about the 
point $q$ there is no actual $q$ dependence, because we 
keep all terms of this expansion. The same is true under the assumption
of vanishing third and higher derivatives: Then the terms up to second
order already make up the full Taylor expansion and thus
the $q$ dependence is only apparent which explains why 
$\Gamma^{(1)}_{\mathrm{res.}}$ and $\tilde{\Gamma}^{(1)}_{\mathrm{res.}}$
coincide in this case.

\section{Discussion and conclusion} \label{conclusions}
\setcounter{equation}{0}

We have calculated the one-loop contribution to 
the scalar effective action with terms up to quadratic order in the 
Taylor expansion of the potential resummed exactly. 
Using world-line methods we could reproduce an earlier result of 
E.~Masso and F.~Rota \cite{Masso:2001zh} in a straight forward
way. We obtained a representation of the full one-loop effective action
with a simple structure (eq.~(\ref{eq:Full1Lp:0-0})): the full expression 
consists of the resummed part containing the total contribution of terms 
up to second order and of a remainder containing the contribution of
higher derivatives. This remainder can easily be taken into account 
to any order desired by calculating Wick contractions of the 
world-line fields with the appropriate Green's function. 

Using a different parametrization of the world-line paths we deduced
an alternative representation of the full one-loop effective action
in terms of a resummed contribution and a contribution of Wick contracted 
world-line fields (eq.~(\ref{eq:Full1Lp:CM})).

If third and higher derivatives of the potential can be neglected
both resummed contributions coincide (and are equal to the full one-loop 
effective action, of course). However, the expression obtained via the
second approach, eq.~(\ref{eq:EffLag:CM=0}), has the advantage of being
much simpler.

If third and higher derivatives can not be neglected, the two resummed
expressions, eqs.~(\ref{eq:EffLag:0-0}) and (\ref{eq:EffLag:CM=0}) are 
no longer equivalent, though the combinations of each resummed part with
the contribution coming from the respective Wick contractions are.
Thus, to calculate the one-loop effective action to a given order
one can either take the resummed contribution and Wick contractions
of the first approach or of the second. Again, the second alternative
seems preferable, because the resummed contribution as well as the Green's
function is simpler in this case.

Finally, let us mention that the method presented here can 
equally well be applied to the case of gauge fields. In the abelian 
case for instance, starting from the world-line representation of the 
one-loop effective action in Fock--Schwinger gauge 
(see ref.~\cite{Fliegner:1997rk})
\begin{eqnarray} \label{eq:Gauge-EffAct}
\lefteqn{
\Gamma^{(1)} [A] 
 =  \int\limits_{0}^{\infty} \frac{dT}{T}\,e^{-m^{2}T} 
    \int\!d^D\!q} \\
& & \times\ \;
    \mathcal{N}^{\,\prime} \!\!\!\!\!
    \oint\limits_{(\mathrm{CM}=0)} \!\!\!\!\mathcal{D}y\,
    \exp\Biggl\{ -\int\limits_{0}^{T}\!\! d\tau 
    \biggl[ \frac{1}{4}\,\dot{y}^{2}(\tau)  +
    ie\!\int\limits_0^1\!\!d\eta\,
    \eta\,y_{\mu}(\tau)
    \,\mathcal{F}_{\mu\nu}\!\left(q + \eta\,y(\tau)\right)
    \dot{y}_{\nu}(\tau)
    \biggr] \Biggr\}
    \nonumber
\end{eqnarray}
and inserting for the field strength tensor its gauge covariant 
expansion (see, e.g.~\cite{Shifman:ui})
\begin{equation}
  \mathcal{F}_{\mu\nu}\!\left(q + \eta\,y(\tau)\right)
  = e^{\eta\,y(\tau) \cdot D}\, \mathcal{F}_{\mu\nu}(q)
\end{equation}
one obtains
\begin{eqnarray} \label{eq:Gauge-EffAct/Expansion}
\lefteqn{
\Gamma^{(1)} [A] 
 =  \int\limits_{0}^{\infty} \frac{dT}{T}\,e^{-m^{2}T} 
    \int\!d^D\!q} \\
& & \times\ \;
    \mathcal{N}^{\,\prime} \!\!\!\!\!
    \oint\limits_{(\mathrm{CM}=0)} \!\!\!\!\mathcal{D}y\,
    \exp\Biggl\{ -\int\limits_{0}^{T}\!\! d\tau 
    \biggl[
      -\frac{1}{2}\,y_{\mu}(\tau)
      \left( \frac{1}{2}\,\delta_{\mu\nu}\partial_{\tau}^{2}  
             -ie\mathcal{F}_{\mu\nu}(q)\,\partial_{\tau}\!
      \right)       y_{\nu}(\tau) 
    \;+\; \V_3(q,y(\tau))
    \biggr]
    \Biggr\}
    \nonumber
\end{eqnarray}
Here we have extracted the first term of the expansion which in the case
of a gauge field is already quadratic in the world-line field $y(\tau)$
and therefore is the contribution that can be resummed exactly. The 
remaining terms of the expansion are collected in
\begin{equation}
  \V_3(q,y(\tau)) = \sum_{n=1}^{\infty}
  \frac{ie}{n!\,(n+2)}\, y_{\mu} (y \cdot D)^n 
  \,\mathcal{F}_{\mu\nu}(q) \,\dot{y}_{\nu}
\end{equation}
as before. If the field strength is constant, $\V_3(q,y(\tau))$
vanishs and thus the resummed effective action can immediately
be read off from the corresponding Euler--Heisenberg action 
\cite{heisenberg} which is a well-known object within the
world-line formalism \cite{Reuter:1996zm,Kors:1998ew,YangMillsEulerH}.
With the result for the one-loop case in 
refs.~\cite{Schmidt:rk,Reuter:1996zm} we find
\begin{equation}
  \Gamma^{(1)}_{\mathrm{res.}}[A] = 
  \int\!d^D\!q\;\,\mathcal{L}^{(1)}_{\mathrm{res.}}[A] =
  \int\!d^D\!q \int\limits_0^{\infty}\!dT\;
  \mathcal{M}^{(1)}_{\mathrm{res.}}[A]
\end{equation}
with
\begin{equation}
  \mathcal{M}^{(1)}_{\mathrm{res.}}[A]  =  
  \frac{1}{(4\pi)^{D/2}} \;
  \frac{e^{-m^2 T}}{T^{1+D/2}} \; 
  \exp\left\{
    -\frac{1}{2}\,\mathrm{tr}\,\ln 
     \frac{\sin e\mathcal{F}(q)T}{e\mathcal{F}(q) T}
  \right\}
\end{equation}
If the assumption of a slowly varying field is not justified, 
the effect of $\V_3(q,y(\tau))$ has to be taken into account.
As we have seen, in the present formalism this can easily be achieved 
by calculating the corresponding Wick contractions of 
the world-line fields by means of the resummed Green's function, i.e.
\begin{equation}
  \Gamma^{(1)}[A] = 
  \int\!d^D\!q \int\limits_0^{\infty}\!dT\;
  \mathcal{M}^{(1)}_{\mathrm{res.}}[A]
  \left\langle
    \exp\biggl\{
      -\int\limits_0^T \!\!d\tau\,
      \V_3\!\left(q,y(\tau)\right)
    \biggr\}
  \right\rangle
\end{equation}
with
\begin{equation}
  \left\langle
    \,y_{\mu}(\tau)\, y_{\nu}(\tau^{\prime})\,
  \right\rangle
  = - \,\mathcal{G}^{(B)}_{\;\mu\nu}(\tau,\tau^{\prime})
\end{equation}
and (see ref.~\cite{Reuter:1996zm})
\begin{equation}
  \mathcal{G}^{(B)}(\tau,\tau^{\prime}) =
  \frac{1}{2(e\mathcal{F})^2}
  \left(
    \frac{e\mathcal{F}}{\sin e\mathcal{F}T}\,
    e^{-ie\mathcal{F}T\dot{G}_B(\tau,\tau^{\prime})}
    + ie\mathcal{F}\dot{G}_B(\tau,\tau^{\prime})
    - \frac{1}{T}
  \right)
\end{equation}
where 
\begin{math}
  \dot{G}_B(\tau,\tau^{\prime}) = \partial_{\tau} G_B(\tau,\tau^{\prime})
\end{math}
and
\begin{math}
  G_B(\tau,\tau^{\prime}) = |\tau-\tau^{\prime}| - (\tau-\tau^{\prime})^2/T
\end{math}.
This then is also a very economised reorganization of the invers mass
expansion in ref.~\cite{Fliegner:1997rk}.

\begin{appendix}
\section{Fixed end points versus fixed center} 
\label{app:EndPointsVersusCenter}
\setcounter{equation}{0}

In this appendix we make a few comments on the relation between world-line
path integrals over the class of all closed paths with a fixed starting and
end point $x(0)=q=x(T)$ and path integrals over the class of all closed paths 
with a fixed ``center of mass''
\begin{equation} \label{def:CenterOfMass}
  q = \frac{1}{T} \int\limits_0^T\!\!d\tau\,x(\tau)
\end{equation}
Originally, the path integrals encountered in the world-line formalism
are of the first kind. The representation of the one-loop effective action,
eq.~(\ref{wpi}), for instance is basically the trace of some operator
evaluated in configuration space 
\begin{equation}
  \textrm{Tr}\,\{O\} = \int\!\!d^D\!q \bra{q} O \ket{q}
\end{equation}
The path integral then arises by introducing complete sets of intermediate
states between $x(0)=q$ and $x(T)=q$. Thus, one typically obtains
a path integral over the set of paths with fixed starting and end point $q$
that is followed by an ordinary integral over all possible positions 
of the point $q$. 
In conjunction, this makes up an integration over all closed 
paths as such, i.e.~without any further restriction.

However, as we have seen in section \ref{second_WL_approach}, sometimes 
it proves useful to parametrize this integration over the total set
of closed paths in a different way: 
defining the ``center of mass'' of a given path 
via eq.~(\ref{def:CenterOfMass}) one first integrates over 
the subset of paths with a fixed ``center of mass'' $q$. 
Afterwards, one integrates all possible positions of
this center. Obviously, this second procedure also comprises an 
integration over the totality of closed paths.

However, one should be aware that depending on the precise definition
of ``integral over all paths with center $q$'' in terms of a 
discretization\footnote{After all, any path integral formula should
always be understood as a mere shorthand notation for a certain
discretization.} there may be a change in the path 
integral measure accompanying the transition from the path integral 
with fixed end points to the path integral with fixed center.

To explain this issue let $F[x]$ be a functional defined 
for paths
$x\!:\![0,T]\rightarrow\mathbbm{R}^D, \tau\mapsto x(\tau)$. We 
choose a discretization where the paths are described by the starting
point $x_0 = x(0)$, the end point $x_{N+1} = x(T)$ and $N$ 
intermediate points $x_1,\dots,x_N$. Then the functional $F[x]$ is 
replaced by an ordinary function of the discretization points
$f(x_0,x_1,\dots,x_N,x_{N+1})$ and we can define the functional
integral as
\begin{equation}
  \int\limits_{x(0)=p}^{x(T)=q} \!\!\!\!\mathcal{D}x\;F[x] =
  \mathcal{N}
  \int
  \biggl[\prod_{m=1}^{N} d^D\!x_m\biggr]
  f(p,x_1,\dots,x_N,q)
\end{equation}
where $\mathcal{N}$ is some given measure. According to this definition
we have
\begin{eqnarray} \label{eq:Int:q-q}
  \int\!\!d^D\!q\!\!\!
  \int\limits_{x(0)=q}^{x(T)=q} \!\!\!\!\mathcal{D}x\;F[x] 
  &=& \mathcal{N}
  \int\!\!d^D\!q
  \int\biggl[\prod_{m=1}^{N} d^D\!x_m\biggr]
  f(q,x_1,\dots,x_N,q) \nonumber \\
  &=& \mathcal{N}
  \int\biggl[\prod_{m=0}^{N} d^D\!x_m\biggr]
  f(x_0,x_1,\dots,x_N,x_0) \label{eq:DiscrQ2Q}
\end{eqnarray}
Analogously, the ``center of mass'' functional translates
\begin{equation} \label{eq:COM}
  C[x] = \frac{1}{T} \int\limits_0^T\!\!d\tau\,x(\tau)
  \quad\longrightarrow\quad
  c(x_0,\dots,x_N) = \frac{\scriptstyle 1}{\scriptstyle N+1} \sum_{k=0}^N x_k
\end{equation}
and can be brought into the game by inserting unity into the integral
(\ref{eq:DiscrQ2Q})
\begin{equation}
  \int\!\!d^D\!q\!\!\!
  \int\limits_{x(0)=q}^{x(T)=q} \!\!\!\!\mathcal{D}x\;F[x] 
  = \mathcal{N}
  \int\biggl[\prod_{m=0}^{N} d^D\!x_m\biggr]
  \int\!\!d^D\!c\;
  \delta^D\big(c- \frac{\scriptstyle 1}{\scriptstyle N+1} 
  \sum_{k=0}^N x_k\big)\;
  f(x_0,x_1,\dots,x_N,x_0)  
\end{equation}
Pulling out the factor of $1/(N+1)$ and using the delta function 
to perform the integral over one of the intermediate points,
say $x_N$, one is led to 
\begin{equation} \label{eq:StartFromFixedEndpoints}
  \int\!\!d^D\!q\!\!\!
  \int\limits_{x(0)=q}^{x(T)=q} \!\!\!\!\mathcal{D}x\;F[x] 
  = 
  \int\!\!d^D\!c\;
  \mathcal{N}^{\,\prime}\,\mathcal{N}\!\!
  \int\biggl[\prod_{m=0}^{N-1} d^D\!x_m\biggr]
  f(x_0,x_1,\dots,x_N(c,x_0,\dots,x_{N-1}),x_0)  
\end{equation}
with $\mathcal{N}^{\,\prime} = (N+1)^D$ and
\begin{math}
  x_N(c,x_0,\dots,x_{N-1}) = (N+1)\,c - \sum_{k=0}^{N-1} x_k
\end{math}.
Note that the factor $(N+1)^D$ is not a combinatorical factor
related to some symmetry of interchanging discretization points.
It is simply a consequence of the specific choice of the restriction
$c=c(x_0,\dots,x_N)$ that the paths are subject to and would be 
different for other choices than (\ref{eq:COM}).

To proceed, let us discuss the expression
\begin{equation} \label{eq:DiscrFixedCenter}
  \mathcal{N}\!\!
  \int\biggl[\prod_{m=0}^{N-1} d^D\!x_m\biggr]
  f(x_0,x_1,\dots,x_N(c,x_0,\dots,x_{N-1}),x_0)  
\end{equation}
occurring on the right-hand side of eq.~(\ref{eq:StartFromFixedEndpoints}). 
It contains $N$ independent integrations over the discretization points
$x_0,\dots,x_{N-1}$. Depending on these values, $x_N$ is choosen in just 
the right way to give the whole path a ``center of mass'' equal to 
the given value $c$. Finally, $x_{N+1}$ is set to $x_0$ constructing a 
closed path. 

Thus, expression (\ref{eq:DiscrFixedCenter}) is exactly what one would 
naturally write down as a discretization of an integral over all 
closed paths with a fixed ``center of mass'' equal to $c$.
Consequently, eq.~(\ref{eq:StartFromFixedEndpoints}) may be written
\begin{equation} \label{eq:EndVersusCenter}
  \int\!\!d^D\!q\!\!\!
  \int\limits_{x(0)=q}^{x(T)=q} \!\!\!\!\mathcal{D}x\;F[x] 
  \ =\  
  \mathcal{N}^{\,\prime}\!
  \int\!\!d^D\!c\!\!\!
  \oint\limits_{(\mathrm{CM}=c)} \!\!\!\!\mathcal{D}x\;F[x]
\end{equation}
and we have shown that the two path integrals are equivalent up to a 
change in the measure.

However, let us argue now that the factor $\mathcal{N}^{\prime}$ 
coming into existence in the transition from path integrals with
fixed end points to integrals with fixed ``center of mass'' will
in general disappear in the end of the calculation and therefore
does not require much attention.

As we have mentioned above a world-line calculation typically starts
from a path integral with fixed end points. If it is transformed
into an integral with fixed center, a factor $\mathcal{N}^{\prime}$
occurs. However, after some manipulations the path integral
will generally end up in an expression proportional to the free
integral over closed paths with ``center of mass'' zero. This free 
integral finally grabs the factor $\mathcal{N}^{\prime}$ to transform 
back into a free integral with fixed end points which is the one we
know to evaluate to $(4\pi T)^{-D/2}$, i.e.
\begin{eqnarray}
  \int\!\!d^D\!q\!\!\!
  \int\limits_{x(0)=q}^{x(T)=q} \!\!\!\!\mathcal{D}x\;F[x] 
  &=&  
  \mathcal{N}^{\,\prime}\!
  \int\!\!d^D\!c\!\!\!
  \oint\limits_{(\mathrm{CM}=c)} \!\!\!\!\mathcal{D}x\;F[x]
  \\
  &=&
  \mathcal{N}^{\,\prime}\!\!\!\!
  \oint\limits_{(\mathrm{CM}=0)} \!\!\!\!\mathcal{D}y\;
  \exp\biggl\{
      -\int\limits_0^T \!d\tau\,
      \frac{1}{4}\,\dot{y}^2
  \biggr\}
  \int\!\!d^D\!c\;(\textrm{some function of }c)
  \nonumber\\
  &=&
  \underbrace{
  \int\limits_{y(0)=0}^{y(T)=0} \!\!\!\!\mathcal{D}y\;
  \exp\biggl\{
      -\int\limits_0^T \!d\tau\,
      \frac{1}{4}\,\dot{y}^2
  \biggr\}
  }_{(4\pi T)^{-D/2}}
  \int\!\!d^D\!c\;(\textrm{some function of }c)
  \nonumber
\end{eqnarray}
where we have used
\begin{equation}
  \int\limits_{y(0)=0}^{y(T)=0} \!\!\!\!\mathcal{D}y\;
  \exp\biggl\{
      -\int\limits_0^T \!d\tau\,
      \frac{1}{4}\,\dot{y}^2
  \biggr\}
  \ =\ 
  \mathcal{N}^{\,\prime}\!\!\!\!
  \oint\limits_{(\mathrm{CM}=0)} \!\!\!\!\mathcal{D}y\;
  \exp\biggl\{
      -\int\limits_0^T \!d\tau\,
      \frac{1}{4}\,\dot{y}^2
  \biggr\}
\end{equation}
which is a consequence of eq.~(\ref{eq:EndVersusCenter}) applied
to
\begin{math}
  F[x]= 
  \exp \{
    -\int_0^T \!d\tau\,
     \frac{1}{4}\,\dot{x}^2
  \}
\end{math}
and a shift of the integration paths in each of the two integrals
according to $y(\tau)=x(\tau)-q$ and $y(\tau)=x(\tau)-c$
respectively.

\end{appendix}


\end{document}